\documentclass[fleqn,usenatbib, letters]{mnras}

\usepackage{newtxtext,newtxmath}
\usepackage{subfigure}

\usepackage[T1]{fontenc}

\DeclareRobustCommand{\VAN}[3]{#2}
\let\VANthebibliography\thebibliography
\def\thebibliography{\DeclareRobustCommand{\VAN}[3]{##3}\VANthebibliography}
\newcommand{\alphaox}{{$\alpha_{\rm ox}$\,}}
\newcommand{\uveddfrac}{{f${_{\rm Edd, UV}}$\,}}
\newcommand{\xeddfrac}{{f${_{\rm Edd, X}}$\,}}
\newcommand{\boleddfrac}{{f${_{\rm Edd, bol}}$\,}}

\usepackage{graphicx}	
\usepackage{amsmath}	
\usepackage{amssymb}	

\title[UV--X-ray correlations in X-ray bright TDEs]{Fainter harder brighter softer: a correlation between \alphaox, X-ray spectral state and Eddington ratio in tidal disruption events}

\author[Thomas Wevers]{Thomas Wevers\thanks{E-mail: tw@ast.cam.ac.uk}\vspace{0.5mm}\\
{\rm Institute of Astronomy, University of Cambridge, Madingley Road, CB3 0HA, UK} \\
}%

\date{Accepted 2020 May 19. Received 2020 April 28; in original form 2020 March 31.}

\pubyear{2020}

\begin{document}
\label{firstpage}
\pagerange{\pageref{firstpage}--\pageref{lastpage}}
\maketitle

\begin{abstract}
We explore the accretion states of tidal disruption events (TDEs) using a sample of 7 X-ray bright sources. To this end, we estimate the relative contribution of the disk and corona to the observed X-ray emission through spectral modeling, and assess the X-ray brightness (through \alphaox, L$_{2\ \rm keV}$ and \xeddfrac) as a function of the Eddington ratio. We report strong positive correlations between \alphaox and \boleddfrac; \xeddfrac and \uveddfrac; and an anti-correlation for L$_{2\ \rm keV}$ and \uveddfrac. TDEs at high \boleddfrac have thermal dominated X-ray spectra and high (soft) \alphaox, whereas those at low \boleddfrac show a significant power-law contribution and low (hard) \alphaox. Similar to X-ray binaries and active galactic nuclei, the transition between X-ray spectral states occurs around \boleddfrac $\approx$ 0.03, although the uncertainty is large due to the small sample size. 
Our results suggest that X-ray surveys are more likely to discover TDEs at low \boleddfrac, whereas optical surveys are more sensitive to TDEs at high Eddington ratios. The X-ray and optical selected TDEs have different UV and X-ray properties, which should be taken into account when deriving rates, luminosity and black hole mass functions. TDEs around the most massive supermassive black holes are observed in the hard state; this could indicate that TDE evolution is faster around more massive BHs.
\end{abstract}


\section{Introduction}
The geometry through which accretion onto compact objects occurs is thought to consist of an accretion disk, surrounded by a hot plasma, dubbed the corona (see e.g. \citealt{Pringle81} for a review). Such a geometry can successfully explain the broad and varied properties of both accreting stellar mass black holes \citep{Remillard2006} and active galactic nuclei (AGN; e.g. \citealt{Done2012}). Which component is energetically dominant, as well as the detailed structure of the accretion flow, depend principally on the accretion rate normalised to the Eddington rate $\dot{m}$ \citep{Abramowicz13}.
Sources accreting at different $\dot{m}$ are expected to have different disk structures, depending on the dominant cooling mechanism (radiation or advection of heat).
Correlation studies between the X-ray spectral state, the two-point UV--X-ray spectral index \alphaox and luminosity Eddington ratio (as a proxy for $\dot{m}$) have been used to constrain the dominant physical mechanisms regulating the accretion flow (e.g. \citealt{Steffen2006, Sobolewska09}).%

Stars that are disrupted by the gravitational tidal forces of supermassive black holes (SMBHs; \citealt{Hills1975, Rees1988}) are signposts for extreme mass accretion rate variability.
In such events, called tidal disruption events (TDEs), $\dot{m}$ increases from (generally) negligible to potentially super Eddington rates in a matter of $\sim$weeks. Their subsequent evolution is observable for $\sim$months -- years (e.g. \citealt{Auchettl2017, vanvelzen20}). 
After disruption, stream-stream collisions are thought to dissipate orbital angular momentum and energy, leading to the formation of an accretion disk \citep{Rees1988}. Depending on the specific orbital dynamics this process may start after a few stream orbital windings (e.g. \citealt{Bonnerot16}), or it may be inefficient (e.g. \citealt{Shiokawa2015}). Observations of bright X-ray emission close to the UV/optical peak as well as optical spectral signatures suggests that at least in some cases, disk formation occurs quickly after disruption \citep{Wevers19b, Short20}.

The detection of late time UV emission (5--10 years after optical peak) suggests that TDE accretion disks are relatively stable and long-lived configurations \citep{vanvelzen2019}. The UV and X-ray lightcurve of the TDE ASASSN--14li can be well modeled by a relativistic thin accretion disk at late times, suggesting that the bulk of the luminosity indeed originates from such a structure \citep{Mummery20}. In addition, X-ray spectral variations \citep{Komossa04, Maksym14}, supported by more recent late time X-ray detections \citep{Jonker20}, provide robust evidence that some TDEs undergo accretion state transitions. \citet{Jonker20} also propose that the X-ray bright phase can be delayed with respect to the UV/optical peak, which helps to explain the different L$_{\rm opt}$/L$_X$ ratios between optical and X-ray selected TDEs. Quantifying such differences provides important input when deriving TDE rates, luminosity and BH mass functions.\\

In this Letter, we analyse multi-wavelength data of 7 X-ray bright TDEs (Section \ref{sec:observations}). We present new correlations between their UV and X-ray properties in Section \ref{sec:results}. These correlations can be interpreted as distinct accretion states/geometries, depending on their Eddington ratio, analogous to X-ray binaries and active galactic nuclei. We find that X-ray selected TDEs are more likely to be discovered at low Eddington ratio, while UV/optical selected TDEs are preferentially discovered at high Eddington ratios. We discuss these results and their implications, in particular differences in population properties depending on selection technique, in Section \ref{sec:discussions}. We summarise the main take-away points in Section \ref{sec:conclusions}.

\section{Observations}
\label{sec:observations}
We search the literature for X-ray bright TDEs with sufficiently good quality X-ray data and UV coverage. We find 10 candidates, of which 3 (AT~2018zr, AT~2018hyz, AT~2019dsg) are excluded on the basis of short lightcurves ($<$ 5 data points) and poor data quality. The resulting sample of 7 TDEs is listed in Table \ref{tab:xrayspectra}, and consists of 4 optical and 3 X-ray selected sources. 

\subsection{UV and X-ray observations}
We use {\it Swift} X-ray Telescope (XRT) and UV/Optical Telescope (UVOT) observations to create lightcurves and stacked spectra for each source. 
UV fluxes are measured using the {\it uvotsource} task in \textsc{heasoft} (V6.24), using a 5$^{\prime\prime}$ aperture and correcting for Galactic extinction using the dust map from \citet{Schlafly2011}. Given that TDEs remain significantly UV bright for 5--10 years after UV/optical peak \citep{vanvelzen2019}, we do not correct for host galaxy contamination. We use simultaneous {\it Swift} UV+U band observations to fit a blackbody model to determine the average UV temperatures, assuming a flat temperature prior 4 $<$ log(T) $<$ 4.7.
Distances are based on optical spectroscopic redshifts, assuming H$_0$=70 km s$^{-1}$ Mpc$^{-1}$ and $\Omega_m$=0.3. 

The X-ray count rates are obtained following the steps outlined in \citet{Evans09}, using the online {\it Swift}/XRT pipeline tool. Count rates are converted to X-ray fluxes using conversion factors depending on the X-ray spectral shape (see Table \ref{tab:xrayspectra}). Stacked spectra are created using the same {\it Swift}/XRT pipeline; they are binned to a minimum of 20 counts per bin, appropriate for using $\chi^2$ statistics to assess goodness of fit. In case the X-ray spectrum changes over time (assessed by first investigating stacked spectra in smaller time bins), we derive multiple conversion factors appropriate for each lightcurve segment. 
\begin{table*}
	\centering
	\caption{Overview of the parameters adopted for the analysis. We provide the X-ray spectral model parameters based on the {\it Swift}/XRT stacked spectra, including temperature $kT$, power law index $\Gamma$, the relative contribution of the power law to the total X-ray flux (0.3--10 keV; PL frac) and the X-ray luminosity (0.3--10 keV) at UV peak. Values marked with an asterisk are frozen to best-fit values obtained from {\it XMM-Newton} observations. References are listed below the table. We also provide the assumed distance (D$_{\rm lum}$, measured velocity dispersion $\sigma$, derived M$_{\rm BH}$, and the peak UV luminosity, the average (bolometric) Eddington ratio and average \alphaox. Multiple entries indicate multiple epochs with significantly different properties (see Table \ref{tab:xrayrange}).}
	\label{tab:xrayspectra}
	\begin{tabular}{lcccccccccccc} %
		Source  & $kT$ & $\Gamma$ & PL frac & D$_{\rm lum}$ & L$_{X}$ &  $\sigma$ & M$_{\rm BH}$ & L$_{\rm UV}$ & T$_{\rm BB}$ & $\langle$\boleddfrac$\rangle$ & $\langle$\alphaox$\rangle$\\
		&eV &&&Mpc&10$^{43}$ erg s$^{-1}$& km s$^{-1}$ & log$_{10}$(M$_{\odot}$) & 10$^{43}$ erg s$^{-1}$ & log$_{10}$(K)& \\
		\hline\hline\vspace{1mm}
		ASASSN--15oi$^{a}$ & 47$^*$ & 2.5$^*$ & 0.07$^{+0.25}_{-0.07}$ & 215 & 0.5$\pm$0.2 &  61$\pm$7 & 5.93$\pm$0.60 & 10.3$\pm$0.3 & 4.45$^{a}$ &  --0.03 & 2.09\\\vspace{1mm}
		  & 42$^*$ & 3.3$^*$ & 0.02 & & &   & & && 0.20 & 2.31\\\vspace{1mm}
		AT~2019azh$^{b}$ & 59$\pm$11 & 2.6$\pm$1.1 & 0.20$\pm$0.20 &96 & 0.08$\pm$0.05 & 77$\pm$2 & 6.36$\pm$0.43 & 28.0$\pm$0.5 & 4.51$^{h}$ & --0.24 & 2.54 \\\vspace{1mm} %
		 & 55$\pm$3 & 5.9$\pm$1.2 & 0.29$^{+0.16}_{-0.25}$ && &  & & && --0.57 & 2.75 \\\vspace{1mm}
		AT~2019ehz$^{c}$ & 138$^{+15}_{-12}$ & 2.8$\pm$1 & 0.05$^{+0.3}_{-0.05}$ &335 &6.2$\pm$0.4 & --- & 6.10$\pm$0.50 &9.7$\pm$0.5& 4.34$^{h}$ & --0.50 & 2.09\\\vspace{1mm}
		AT~2018fyk$^{d}$ & 142$\pm$10 & 2.8$\pm$1 & 0.25$\pm$0.13 & 264 & 13.3$\pm$0.7 & 158$\pm$1 & 7.69$\pm$0.39 & 48$\pm$2 & 4.63 & --1.11 & 1.91  \\\vspace{1mm}
		 & 179$\pm$8 & 2.20$\pm$0.08 & 0.68$\pm$0.04 && & & & && --1.89 & 1.22 \\\vspace{1mm}
		XMMSL1J0740$^{e}$ & 207$\pm$27 & 1.95$\pm$0.29 & 0.59$\pm$0.11&75 &0.48$\pm$0.03 & 112$\pm$3 & 7.05$\pm$0.43 & 0.62$\pm$0.01 & 4.24 & --2.33 & 1.55 \\\vspace{1mm}
		XMMSL2J1446$^{f}$ & 113$^*$ & 2.58$^*$ & 0.87$^*$ &127 & 0.37$\pm$0.05  & 167$\pm$15 & 7.79$\pm$0.55 & 0.33$\pm$0.01 & 4.15 & --2.61 & 1.44\\\vspace{1mm}
		XMMSL1J1404$^{g}$ & 123$\pm$10 & 2.7$\pm$0.33 & 0.38$^{+0.13}_{-0.09}$ &190 &3.0$\pm$0.3  &93$\pm$1 & 6.71$\pm$0.40 &0.29$\pm$0.05 & 4.22 &--0.90 & 1.25
	\end{tabular}
	    \begin{flushleft}
    \small $^{a}$\citet{Holoien201615oi,Gezari17}, $^{b}$aka ASASSN--19dj, \citet{Liu19}, $^{c}$aka Gaia19bpt, \citet{vanvelzen20}, $^{d}$aka ASASSN--18ul, \citet{Wevers19b}, $^{e}$ \citet{Saxton17} $^{f}$\citet{Saxton19}, $^{g}$Saxton et al., in prep., $^{h}$ \citet{vanvelzen20}
    \end{flushleft}
    
\end{table*}

To estimate the relative strength of the thermal and power-law components in the X-ray spectra, we use a phenomenological 2-component spectral model in \textsc{Xspec}: TBabs $\times$ zashift $\times$ cflux $\times$ (diskbb + powerlaw). The hydrogen column density is fixed to the Galactic foreground values from \citet{hi4pi}, and uncertainties are computed using the \textsc{error} command. Our goal is not to provide the absolute best fit for each individual object (i.e. a different model may have a lower reduced $\chi^2$, $\chi_r$), but to adopt a systematic approach to ensure simplicity as well as allowing a direct comparison between the quantities of interest across our sample. We do note that the fits are generally statistically acceptable ($\chi_r < 1.3$), although some best fit models have $1.3 < \chi_r < 1.4$.

After fitting the stacked {\it Swift}/XRT spectra, we convert the observed countrates to fluxes (0.3--10 keV) using the best fit spectral parameters. We then use the best fit spectral model to assess the fractional contribution of each spectral component to the total flux; the power law contribution is subsequently used to estimate the monochromatic 2 keV X-ray flux (L$_{2\ \rm keV}$), using the best fit indices from either the {\it Swift}/XRT stacked spectra or from the literature (if based on better quality {\it XMM-Newton} spectra; Table \ref{tab:xrayspectra}). Given the low blackbody temperatures, the power law component dominates the X-ray flux at 2 keV at all times. Uncertainties are propagated using standard conventions. 

Finally, we calculate the UV to X-ray spectral slope \alphaox following \citet{Tananbaum79}
\begin{equation}
    \alpha_{ox} = 1 - \frac{log_{10}(\lambda L_{2500}) - log_{10}(\lambda L_{2 keV})}{log_{10}(\nu_{2500}) - log_{10}(\nu_{2 keV})}
\end{equation}
We use the $UVW1$ fluxes ($\lambda_{cen}$ = 2629 \AA) as a proxy for L$_{2500}$. 

\subsection{Black hole masses and Eddington ratios}
\label{sec:opticalspectra}
Optical spectroscopic observations were obtained with the William Herschel (ISIS) and Magellan (MagE) telescopes (details can be found in Table \ref{tab:opticalspectra}), and reduced using \textsc{iraf} or the CarPY software \citep{Kelson03}.
All spectra are normalised to the continuum by fitting a low order spline function while masking absorption and emission lines.
Following the procedure outlined in \citet{Wevers17}, we use the Penalised Pixel Fitting routine \textsc{ppxf} to measure the velocity dispersion of the spectra \citep{Cappellari17}. 

We take these velocity dispersion measurements, as well as values from \citet{Wevers19a} for ASASSN--15oi and XMMSL2J1446, and use the M\,--$\sigma$ relation of \citet{Gultekin09} (including both early and late type galaxies) to convert them to black hole masses. Uncertainties are obtained by linearly combining the intrinsic scatter in the relation with the statistical (measurement) uncertainties in the velocity dispersions. 
For AT~2019ehz, we use SDSS and Pan-STARRS1 PSF and Petrosian/Kron fluxes to estimate a bulge-to-total correction of (B/T)$_g$ = 0.23. We combine this with the galaxy stellar mass from \citet{vanvelzen20} and the M$_{\rm BH}$\,--\,M$_{\rm bulge}$ relation of \citet{Haring04} to estimate the black hole mass. Including an intrinsic scatter of 0.3 dex and a factor 0.2 dex uncertainty in the bulge mass, this yields log$_{10}$(M$_{\rm BH}$) = 6.1$^{+0.5}_{-0.5}$ M$_{\odot}$.

The Eddington luminosity is calculated as L$_{\rm Edd} = 1.26\ \times\, 10^{38}\, M_{\rm BH}\ {\rm erg\ s^{-1}}$. To account for the fact that the UVW1 filter is on the Rayleigh-Jeans tail of the blackbody emission (Table \ref{tab:xrayspectra}), we apply a temperature correction (see \citealt{vanvelzen2019}) to obtain the UV luminosity L$_{\rm UV}$ (not to be confused with L$_{UVW1}$), and define Eddington ratios \uveddfrac\,$\equiv$\,L$_{\rm UV}$\,/\,L$_{\rm Edd}$ and \xeddfrac\,$\equiv$\,L$_{X}$\,/\,L$_{\rm Edd}$. To account for the potentially important contribution of the X-ray emission at EUV wavelengths, 
we find the bolometric luminosity by extrapolating the X-ray spectral model into the energy range 0.01--10 keV and adding L$_{\rm UV}$, such that \boleddfrac \,$\equiv$\,(L$_{0.01-10\ \rm keV}$+L$_{\rm UV}$)\,/\,L$_{\rm Edd}$.
Uncertainties in f$_{\rm Edd}$ include both the black hole mass and photometric measurement uncertainties, with the former being dominant.

\section{Analysis and results}
\label{sec:results}
\subsection{Correlation analysis}
Power law correlations between several variables such as \alphaox, L$_{2\ \rm keV}$, $\Gamma$ and Eddington ratio have been found in the active SMBH population \citep{Steffen2006, Vasudevan07, Lusso10}. Therefore, we investigate possible correlations between observables in log space, using a hierarchical Bayesian approach to linear regression (\textsc{linmix}). This method allows to take into account heteroscedastic measurement errors and covariances \citep{Kelly2007}. 
\begin{figure*}
	\begin{subfigure}
	\centering
	\includegraphics[width=\linewidth]{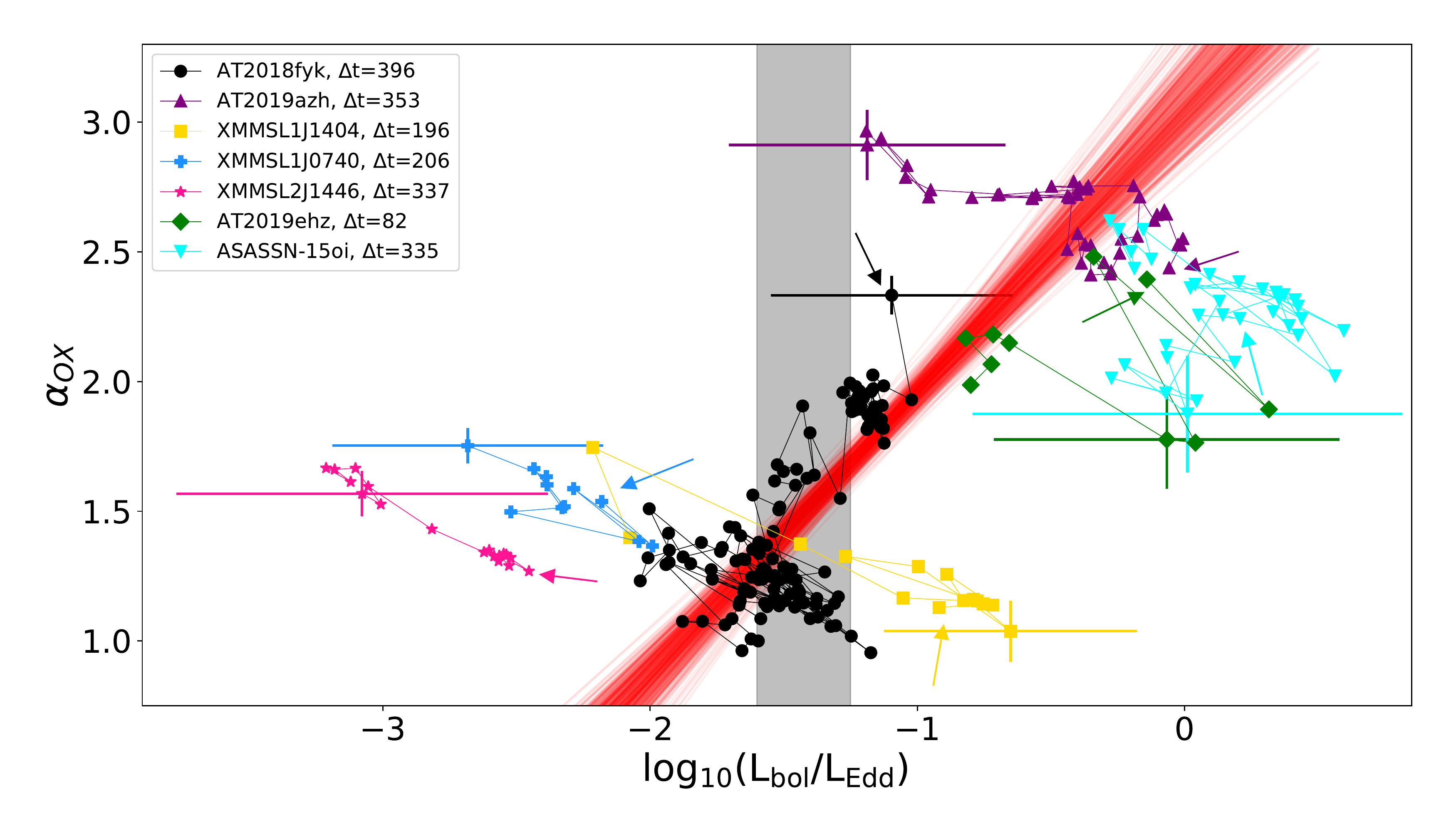}
    \end{subfigure}
    \begin{subfigure}
    \centering
    \includegraphics[width=0.49\linewidth]{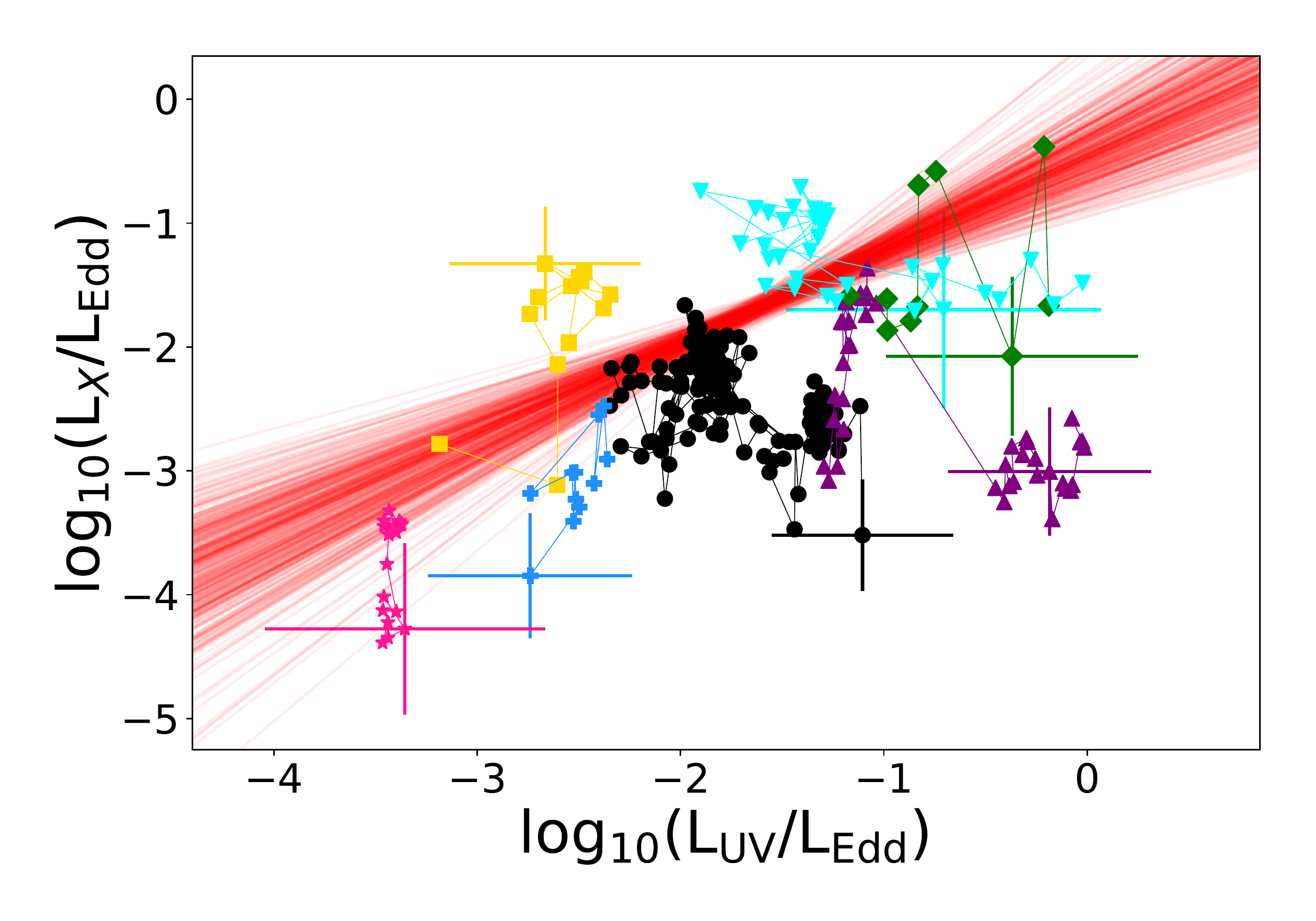}
    \end{subfigure}
    \hfill
    \begin{subfigure}
    \centering
    \includegraphics[width=0.49\linewidth]{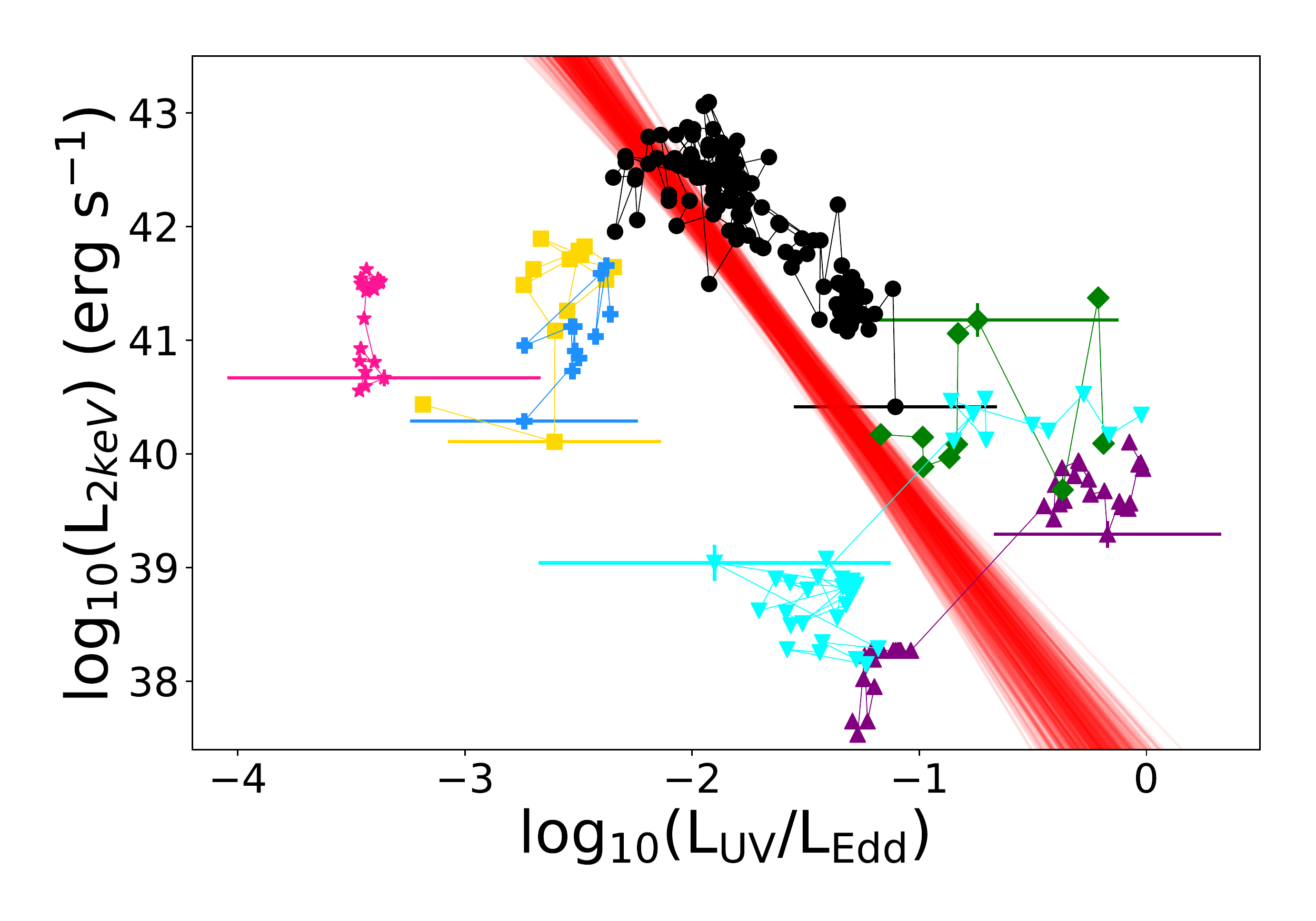}
    \end{subfigure}
    \caption{Top: UV to X-ray spectral slope ($\alpha_{ox}$) as a function of \boleddfrac. Red lines are samples drawn from the posterior distributions of the best fit regression line. The black shaded region indicates the approximate division between disk-dominated (higher f$_{\rm Edd}$) and power-law dominated (lower f$_{\rm Edd}$) X-ray spectra. The first observation is marked by a coloured arrow to indicate the time evolution; $\Delta$t is the lightcurve duration, in days. We show the median error bar for each source. 
    Bottom left: best fit regression line between \xeddfrac and \uveddfrac. Bottom right: best fit regression line between L$_{2\ \rm keV}$ and \uveddfrac. }
    \label{fig:uvfedd_alphaox}
\end{figure*}

We report a strong, statistically significant correlation between \alphaox and \boleddfrac. The best fit regression line (Figure \ref{fig:uvfedd_alphaox}) is given by \begin{equation}
\alpha_{ox} = (3.03\pm0.11) + (1.08\pm0.08) \ {\rm log}_{10}({\rm f}_{\rm Edd, bol})
\end{equation}
quoting posterior distribution median estimates, and the uncertainties are the standard deviations of the posteriors. 
The correlation coefficient is constrained to $\hat{\rho} > 0.95$. 
The effect of systematic uncertainties in the M--$\sigma$ relation is assessed by sampling each M$_{\rm BH}$ within its uncertainty, and performing the linear regression using the resulting \boleddfrac values. Using 200 trials, we find identical results, but with a scatter $\sim$5 times larger than the statistical uncertainties.

The observed \alphaox -- \boleddfrac correlation is likely driven by a strong negative correlation between L$_{2\ \rm keV}$ and \uveddfrac, such that the 2 keV X-ray emission is weaker when \uveddfrac is higher (Fig. \ref{fig:uvfedd_alphaox}, bottom right panel):
\begin{equation}
{\rm L}_{2 keV} = (37.0\pm0.3) - (2.58\pm0.18) \ {\rm log}_{10}({\rm f}_{\rm Edd, UV})
\end{equation}
The correlation coefficient is constrained to $\hat{\rho}$=--0.87\,$\pm$\,0.02.

Finally, we report a correlation between \xeddfrac and \uveddfrac, with $\hat{\rho}$=0.78$\pm$0.10 (Figure \ref{fig:uvfedd_alphaox}, lower left panel):
\begin{equation}
{\rm f}_{\rm Edd, X} = (-0.43\pm0.23) + (0.78\pm0.13) \ {\rm log}_{10}({\rm f}_{\rm Edd, UV})
\end{equation}

These correlations remain statistically significant (10, 10 and 6 $\sigma$ respectively) when removing AT~2018fyk from the sample.

We note that we also recover the canonical L$_{\rm UV}$--L$_{2 keV}$ relation seen in AGN \citep{Lusso16}, although the correlation coefficient is weak ($\hat{\rho}$ = 0.34$\pm$0.05). We find $\hat{\alpha}$= 10.6$\pm$5.2, $\hat{\beta}$= 0.70$\pm$0.12, fully consistent with the \citet{Lusso16} relation -- albeit with much larger scatter in the parameters due to the small sample size.

\subsection{X-ray spectral state as a function of \boleddfrac}
The results of our X-ray spectral fitting are provided in Table \ref{tab:xrayspectra}. We find that broadly speaking, the TDEs in our sample can be divided into blackbody or power law dominated spectral states. The former sources have on average higher \alphaox values (2.28$\pm$0.32), lower power law fractions (0.17$\pm$0.12), as well as somewhat steeper power law indices, and are found at \boleddfrac $\gtrsim 0.03$. The power law dominated sources have lower \alphaox values (1.33$\pm$0.17), higher power law fractions (0.63$\pm$0.18), shallower power law indices and are found at \boleddfrac $\lesssim$ 0.03. This transition value of \boleddfrac is subject to considerable uncertainties, dominated by the intrinsic scatter in the M--$\sigma$ relation. 

The average black hole masses for the blackbody and power law dominated spectral states are mutually compatible within the typical uncertainty, although the sample size is small and thence the scatter large. 
\section{Discussion}
\label{sec:discussions}
\subsection{Potential contamination of \alphaox}
\label{sec:contamination}
\citet{Mummery20} showed that at very early times there may be significant contamination from an additional, exponentially decaying energy source at UV wavelengths (in excess of the disk emission). This emission decays in $\approx$200 days for ASASSN--14li. Similar behaviour likely occurs for most UV/optical discovered TDEs, given their similar lightcurve evolution and properties (e.g. \citealt{Hinkle20, vanvelzen20}). 

As a result, our \alphaox values at early times may be similarly contaminated due to UV emission unrelated to the disk. However, Figure \ref{fig:uvfedd_alphaox} (top panel) shows that even the {\it late time} emission in ASASSN--15oi and AT~2019azh (observed $\sim$250 days after the UV peak, after the relatively large gaps in the \boleddfrac evolution) has a significantly higher \alphaox than the X-ray selected sources. The fact that these late time data are closer to the regression line may indicate that some contamination is indeed present at early times. Nevertheless, their X-ray spectra have similar properties at {\it all} times, and in the case of ASASSN--15oi, narrow Fe\,\textsc{ii} emission indicates that we are observing the disk directly at late times, also in optical light \citep{Wevers19b}. Removing the low X-ray luminosity data (close to UV/optical peak) for ASASSN--15oi and AT~2019azh steepens the L$_{2\ \rm keV}$--\uveddfrac correlation ($\hat{\alpha}$=38.05$\pm$0.25 and $\hat{\beta}$=--2.14$\pm$0.15). 

There is no evidence for systematically higher gas column densities in the X-ray spectra of the high \alphaox sources (e.g. \citealt{Auchettl2017}). Barring extreme gas-to-dust ratios, this also implies that systematically higher dust obscuration in the UV for the low \alphaox sources is disfavoured. We also note that including potential host galaxy contamination in the relatively UV faint X-ray selected TDEs would make the distinction in \alphaox more pronounced, as they would be intrinsically more X-ray strong (i.e. move to lower \alphaox). 

\subsection{Accretion state transitions as a function of \boleddfrac}
\label{sec:transition}
Our analysis shows that sources at higher Eddington ratio have higher \alphaox values, i.e. their 2 keV X-ray luminosity is weaker for their Eddington scaled UV luminosity. In Figure \ref{fig:uvfedd_alphaox} we have provided a tentative shaded region to guide the eye; sources to the right of this region (around \boleddfrac $\approx$ 0.03) all have X-ray spectra dominated by a thermal component (Table \ref{tab:xrayspectra}). Sources on the left of this region have an X-ray spectrum dominated by a power law.

In AGN there is evidence for correlations similar to those presented here \citep{Steffen2006, Vasudevan07, Lusso10}. Moreover, marked changes in X-ray spectral state and SED appear to coincide with sources crossing a {\it transition} value of \boleddfrac $\sim 0.02$ in both AGN and XRBs \citep{Maccarone03, Remillard2006}, although with large scatter in some cases \citep{Maccarone2003}. We find, encouragingly, that the transition occurs at very similar values in TDEs.
While several attempts to explain the observed correlations with simple models for AGN accretion flows exist \citep{Kubota18, Arcodia19}, it remains unclear what the dominant physical mechanism to explain the interplay between the disk and corona is.

The strong negative correlation observed between L$_{2\ \rm keV}$ and \uveddfrac shows that a mechanism regulating the energetic interaction between the corona and UV photon field must exist. Although other sources of UV radiation (e.g. the exponentially declining component reported in \citealt{Mummery20}) may also play a role, we note that the UV blackbody component at peak in TDEs is produced at radii much larger than the putative accretion disk \citep{Wevers19a}. Systematic uncertainties in M$_{\rm BH}$ also do not appear to explain the observed behaviour. We do note that the most massive sources are all observed in the hard state, which may indicate that their evolution is faster than TDEs around low mass black holes.

The apparent flattening of \alphaox at very high and very low \boleddfrac may instead reflect changes in disk structure and/or dominant emission mechanism. At low \boleddfrac, a transition from a thin disk to an advection dominated accretion flow will lead to softer emission, potentially through a shift in dominant X-ray emission mechanism, or changes in radiative transfer in the disk photosphere (see e.g. \citealt{Ruan19a}, and references there-in). There is some evidence of this process occurring in XMMSL2J1446 (Fig. \ref{fig:uvfedd_alphaox}). At high \boleddfrac, a transition to a slim disk may prevent arbitrarily soft \alphaox due to an increased scale height and/or advective cooling. 

\subsection{Potential selection effects in X-ray surveys}
Given that the properties of both the X-ray and optically selected TDEs are consistent with the state transition scenario, it seems unlikely that the X-ray selected TDEs are probing an intrinsically different class of objects. Moreover, there are no significant differences between X-ray and optically selected TDE host black hole masses \citep{Wevers19a}, although the sample size prevents robust conclusions \citep{French20}. The low observed peak UV luminosities of the X-ray selected TDEs open the possibility that the observed X-ray emission is significantly delayed with respect to the UV/optical peak of the lightcurve. This scenario was recently proposed by \citet{Jonker20} to explain the differences in L$_{\rm opt}$/L$_X$ of optical and X-ray selected TDEs. 

A physically motivated explanation could be a viscous delay between disruption and the onset of accretion, where material is unable to circularise efficiently, delaying the emergence of accretion powered emission \citep{Guillochon2015}. This scenario was invoked by \citet{Gezari17} to explain the behaviour of ASASSN--15oi, and later also for AT~2019azh \citep{Liu19}. In these sources, the X-ray emission is observed to increase over a period of $\sim$250 days, while the UV emission fades. These two sources were discovered at optical wavelengths close to UV/optical peak, and are therefore observed at high \boleddfrac. The properties of the X-ray selected TDEs are compatible with a similar evolution, provided that we missed the peak of the UV/optical emission. On the other hand, they do show significant evolution over time \citep{Saxton17}, suggesting that the gap is likely not more than about a year. An alternative scenario with a time-variable absorbing column density (higher at early times due e.g. to an increased scale height of the inner disk or a wind), can also explain the observed properties \citep{Wen20}.

The X-ray selected sources in our sample appear to have been detected when they were very X-ray strong (low \alphaox). Taking into consideration the delayed accretion scenario as well as the discovered correlations with \boleddfrac and \uveddfrac, this will result in X-ray surveys preferentially detecting TDEs at lower Eddington ratios and \alphaox, well after the UV/optical peak. As a caveat we note that the selection function also depends on survey cadence and duration of the X-ray peak of the lightcurve, which is not well constrained at present.
Nevertheless, a sample of X-ray bright, optical selected TDEs combined with X-ray selected TDEs will cover the full range of Eddington ratios from $\sim$10$^{-4}$ -- 1 (see Figure \ref{fig:uvfedd_alphaox}), which are of interest to study accretion state transitions in SMBHs.

\section{Conclusions}
\label{sec:conclusions}
We summarise our main findings below:
\begin{itemize}
\item Using 7 TDEs we find statistically significant correlations between \alphaox, \xeddfrac and Eddington ratio, as well as an anti-correlation between L$_{2\ \rm keV}$ and \uveddfrac. Using linear regression, we find slopes inconsistent with 0 at the 14, 14 and 6 $\sigma$ level, respectively. The emission significantly softens at higher Eddington ratio; in other words, the power law X-ray emission is suppressed more rapidly than the UV luminosity when moving to higher \boleddfrac.\\

\item We find that on average, sources at higher \boleddfrac have disk dominated X-ray spectra, with high \alphaox values and a small power law contribution to the total X-ray flux. Sources at lower \boleddfrac have power law dominated X-ray spectra and lower \alphaox values. A spectral state transition occurs around \boleddfrac $\sim$0.03, although there is significant uncertainty in this value, due to the small sample size and M$_{\rm BH}$ uncertainties. TDEs around the most massive SMBHs are observed in the hard state; this could indicate that TDE evolution is faster around more massive black holes.\\

\item Our results show that the UV and X-ray properties of optical and X-ray selected TDEs can differ significantly. Such selection effects must be taken into account when computing TDE rates, luminosity and BH mass functions, particularly in the X-ray band.\\

\item The small sample used here covers Eddington ratios spanning 4 orders of magnitude, from 10$^{-4}$ to 1. We find both qualitative and quantitative similarities with XRB and AGN accretion flows. The rapid evolution from Eddington limited to low accretion rates makes TDEs promising tools to study changes in accretion flow structure in individual SMBHs.\\

\item The X-ray emission appears to be delayed with respect to the UV/optical peak for several optical TDEs \citep{Jonker20}. In combination with the reported correlations, this implies that X-ray surveys are more likely to find TDEs in the low Eddington accretion regime, while UV/optical surveys are more sensitive to the high Eddington regime. Combining X-ray and optical samples will provide the best leverage in studying accretion properties over a large range of f$_{\rm Edd}$.\\
\end{itemize}

Large samples of TDEs detected with current and future X-ray and optical surveys will allow statistical studies of the accretion flow formation, properties and evolution (e.g. disk-corona geometry and interplay), providing new insights into the fundamental physics governing accretion in black holes.

\section*{Acknowledgements}
I thank Matt Auger, Peter Jonker, Dom Walton and Martha Irene Saladino for insightful comments and discussions; the referee, Richard Saxton, for astute comments and suggestions in prompt and constructive reports; and DJ Pasham and Ani Chiti for performing the Magellan observations. This work is funded by ERC grant 320360 and European Commission grant 730980. 
We acknowledge the use of public data from the {\it Swift} data archive. This paper includes data gathered with the 6.5 meter Magellan Telescopes located at Las Campanas Observatory, Chile, as well as observations obtained with the William Herschel Telescope (proposal ID: W19B/P7) operated on the island of La Palma by the Isaac Newton Group of Telescopes in the Spanish Observatorio del Roque de los Muchachos of the Instituto de Astrofisica de Canarias.

\bibliographystyle{mnras}
\bibliography{bibliography} 

\appendix 
\section{Observing log and additional information}
\begin{table}
	\centering
	\caption{Observing log of the optical spectroscopy. The slit width was 0.7 arcsec for all observations. The observations of AT~2018fyk were seeing-limited, resulting in a better spectral resolution.}
	\label{tab:opticalspectra}
	\begin{tabular}{lcccc} 
		\hline
		Name & Instr & Exp. time & $\sigma_{instr}$  & Obs. date \\
		&& sec & km s$^{-1}$ &   \\
		\hline\hline
		AT~2018fyk & MagE & 2x1800 &  17 &  2019/08/13 \\
		AT~2019azh & ISIS/R600 &  2700 & 45 & 2020/02/02\\
		XMMSL1J0740 & MagE  & 800 & 22 &  2020/02/15 \\
		XMMSL1J1404 & MagE & 2700 & 22 &  2020/02/15 \\
		\hline
	\end{tabular}
\end{table}

\begin{table}
	\centering
	\caption{Julian date range for each epoch referenced in Table 1.}
	\label{tab:xrayrange}
	\begin{tabular}{lc} 
		\hline
		Name & MJD \\
		& days   \\
		\hline\hline
ASASSN--15oi & 57\,276-57\,340 \\
&57\,468-57\,611 \\
AT2019azh & 58\,553-58\,635 \\
& 58\,767-58\,906 \\
AT2019ehz &58\,620-58\,702 \\
AT2018fyk & 58\,383-58\,447 \\
 & 58\,579-58\,779\\
XMMSL1J0740 & 56\,782-56\,988 \\
XMMSL2J1446 & 57\,643-57\,980 \\
XMMSL1J1404 &58\,177-58\,373\\
		\hline
	\end{tabular}
\end{table}

\label{lastpage}
\end{document}